\documentclass[onecolumn,showpacs]{revtex4}

\usepackage{graphicx}
\usepackage{dcolumn}
\usepackage{bm}

\input epsf

\begin{document}

\title{\Large  THE GEOMETRY OF THE HIGHER DIMENSIONAL BLACK HOLE THERMODYNAMICS IN EINSTEIN-GAUSS-BONNET THEORY
}

\author{\bf Ritabrata
Biswas\footnote{ritabratabiswas@rediffmail.com}~And~Subenoy
Chakraborty.\footnote{schakraborty@math.jdvu.ac.in} }

\affiliation{Department Of Mathematics, Jadavpur University}

\date{\today}

\begin{abstract}

This paper deals with five-dimensional black hole solutions in (a)
Einstein-Yang-Mills-Gauss-Bonnet theory and
(b)Einstein-Maxwell-Gauss-Bonnet theory with a cosmological
constant for spherically symmetric space time. The geometry of the
black hole thermodynamics has been studied for both the black
holes.

\end{abstract}

\pacs{95.36.+x,04.70.Dy, 04.60.Kz.}

\maketitle

\section{INTRODUCTION}

The nice similarity between a black hole and a thermodynamical
system is characterized by temperature and entropy: a black hole
temperature (known as Hawking temperature) is proportional to its
surface gravity on the horizon while entropy proportional to its
horizon area$[1,2]$ and they satisfy the first law of
thermodynamics$[3]$. But the statistical origin of the black hole
entropy is still a challenging problem today. Usually, a black
hole is characterized by three parameters namely its mass, charge
and angular momentum(known as no hair theorem) and the
thermodynamical stability of the black hole is determined by the
sign of its heat capacity$(c_{v})$. If $c_{v}<0$ (as Schwarzschild
black hole), then black hole is thermodynamically unstable, but if
$c_{v}$ changes sign in the parameter space such that it
diverges$[4]$ in between then from ordinary thermodynamics it
indicates a second order phase transition $[5]$. However in black
hole thermodynamics, a critical point exits also for extremal
black hole and the second order phase transition takes place from
an extremal black hole to its non extremal counter part.
    The geometrical concept into ordinary thermodynamics was first
introduced by Weinhold$[6]$(for details see the review$[7]$). According
 to him a Riemannian metric can be defined as the second derivative of
the internal energy($U$) to entropy ($S$) and other extensive variables
 ($y^{\alpha}$) of the system ,i.e., the Hessian of the energy, known as Weinhold metric as

\begin{equation}
g_{i j}^{(W)} = \partial_{i}\partial_{j}U(S,y^{\alpha})
\end{equation}

However it looses physical meaning in equilibrium thermodynamics.
Subsequently, Ruppeiner$[8]$ introduced a metric as the Hessian
matrix of the thermodynamic entropy and is known as the Ruppeiner
metric. It is defined on the state space as

\begin{equation}
g_{i j}^{(R)} = -\partial_{i}\partial_{j}S(U,y^{\alpha})
\end{equation}

Note that the Ruppeiner metric is conformally related to the
Weinhold metric with $T^{-1}$ as the conformal factor ,i.e,
\begin{equation}
ds_{R}^{2} = \frac{1}{T}ds_{W}^{2}
\end{equation}

Unlike Weinhold metric , the Ruppeiner geometry has physical
relevance in the fluctuation theory of equilibrium thermodynamics.
In fact, the inverse Ruppeiner metric gives the second moments of
fluctuations. It was found that the Ruppeiner geometry is related
to phase structure of thermodynamic system such that scalar
curvature of the Ruppeiner metric diverges $(R \rightarrow
-\infty)$ at the phase transition and critical point while the
Ruppeiner metric is flat(i.e scalar curvature is zero) for
thermodynamical system having no statistical mechanical
interactions.

    In the present paper we study the geometry of the black hole
thermodynamics in the five dimensional (a)
Einstein-Yang-Mills-Gauss-Bonnet theory(EYMGB) and
(b)Einstein-Maxwell-Gauss-Bonnet(EMGB) theory with a cosmological
constant. With the recent development in String Theory, it is
legitimate to assume the space time dimension to be more than four
particularly in high energy physics. Also the classical
gravitational analog of String Theory$[9]$demands a generalization
of Einstein -Hilbert action by squares and higher powers of
curvature terms, but field equations become fourth order and bring
in Ghost$[10]$. This problem was resolved by Lovelock$[11]$
choosing higher powered curvature terms in a particular
combination so that the field equations still remain second order
and the ghosts disappear. The first two terms in the Lagrangian of
the Lovelock gravity are the Einstein-Hilbert (EH) and
Gauss-Bonnet(GB)terms and they form the most general Lagrangian
for the second-order field equations in the five dimension$[12]$.
The GB term is consistent with the heterotic Super String Theory
and play a fundamental role in Chern-Simons gravity$[13]$. One may
note that GB term is purely topological in nature$[14]$ in 4-D and
hence it has no role in dynamics. This paper is organized a
follows: section $II-A$ describes the black hole solution in
Einstein-Yang-Mills-Gauss-Bonnet theory and the corresponding
geometrical aspect of black hole thermodynamics have been studied
in section $II-B$ The black hole in Einstein-Maxwell-Gauss-Bonnet
theory is presented in section $III-A$ and its thermodynamics has
been examined from geometric point of view in section $III-B$

\section{THERMODYNAMIC GEOMETRY OF EYMGB BLACK HOLES}

In this section we first describe the black hole solution in
Einstein-Yang-Mills-Gauss-Bonnet (EYMGB)theory and then study the
geometry of the black hole thermodynamics in the next subsection:

\subsection{SPHERICALLY SYMMETRIC YANG MILLS SOLUTION IN EINSTEIN-GAUSS-BONNET THEORY}

Mazharimousavi and Halisoy$[15]$ have recently obtained a 5-D
spherically symmetric solution in EYMGB theory. The metric ansatz
for 5-D spherically symmetric space-time is chosen as
\begin{equation}
ds^{2} = -U(r)dt^{2}+\frac{dr^{2}}{U(r)}+r^{2}d\Omega_{3}^{2}
\end{equation}

where they have expresed the metric on unit three sphere
$d\Omega_{3}^{2}$ in terms of Euler angles as$[15]$
\begin{equation}
d\Omega_{3}^{2}
=\frac{1}{4}(d\theta^{2}+d\phi^{2}+d\psi^{2}-2\cos\theta d\phi
d\psi )
\end{equation}

with ~~~$\theta ~\epsilon~[0,\pi],~~~ (\phi,~\psi)~\epsilon~[0,
2\pi]$.

For the Yang-Mills field the energy momentum tensor is given by,
\begin{equation}
T_{\mu\nu}=2F_{\mu}^{i \alpha}F^{i}_{\nu \alpha}-\frac{1}{2}g_{\mu
\nu}F^{i}_{\alpha \beta}F^{i \alpha \beta}
\end{equation}

where $F^{i}_{\alpha \beta}$ is the Yang-Mills field 2-forms such
that $F^{i}_{\alpha \beta}F^{i \alpha
\beta}=\frac{6Q^{2}}{r^{4}}$, $Q$ the only non-zero gauge charge.
The modified Einstein equations in EYMGB theory are
\begin{equation}
G_{\mu \nu}-\alpha H_{\mu \nu}=T_{\mu\nu}
\end{equation}

Where $G_{\mu \nu}$ is the usual Einstein tensor in 5-D,
$T_{\mu\nu}$ is the energy-momentum tensor given by equation
$(6)$, $\alpha$, the GB coupling parameter is chosen to be
positive in the heterotic string theory and the Lovelock tensor
has the expression
\begin{equation}
H_{\mu \nu}=\frac{1}{2}g_{\mu \nu}(R_{\alpha \beta \gamma
\delta}R^{\alpha \beta \gamma \delta}-4R_{\alpha \beta} R^{\alpha
\beta}+R^{2})-2RR_{\mu \nu}+4R^{\lambda}_{\mu}R_{\lambda
\nu}+4R^{\rho \sigma}R_{\mu \rho \nu \sigma}-R^{\alpha \beta
\gamma}_{\mu}R_{\nu \alpha \beta \gamma}
\end{equation}

Now solving the non vanishing components of the field equations we
have$[15]$
\begin{equation}
U(r)=1+\frac{r^{2}}{4\alpha}\pm\sqrt{\left(\frac{r^{2}}{4\alpha}\right)^{2}+\left(1+\frac{m}{2
\alpha}\right)+\frac{Q^{2} \ln r }{\alpha}}
\end{equation}

with $m$ as the constant of integration. Now as $\alpha
\rightarrow 0$

\begin{equation}
U(r) \rightarrow 1-\frac{m}{r^{2}}-\frac{2 Q^{2} \ln r}{r^{2}}
\end{equation}

provided negative branch is considered. The metric coefficient
$U(r)$ in $(10)$ is identical to that of the Einstein-Yang-Mills
solution and hence $'m'$ is interpreted as the mass of the system.
In the equation $(9)$ for $U$ the expression within the square
root is positive definite for $\alpha>0$ while the geometry has a
curvature singularity at the surface $r=r_{min}$ for $\alpha<0$ .
Here $r_{min}$ is the minimum value of the radial coordinate such
that the function under the square root is positive. Moreover,
depending on the values of the parameters $(m,Q,\alpha)$, the
singular surface can be surrounded by an event horizon with radius
$r_{h}$ so that the space-time given by equation $(1)$ represents
a black hole. However if no event horizon exists, then there will
be naked singularity.

    Now the metric decribed by equation $(1)$ and $(9)$ has a singularity
at the greatest real and positive solution ($r_{s}$)of the
equation
\begin{equation}
\frac{r^{4}}{16 \alpha^{2}}+\left(1+\frac{m}{2
\alpha}\right)+\frac{Q^{2} \ln r}{\alpha}=0
\end{equation}

Note that if equation $(11)$ has no real positive solution then
the metric diverges at $r=0$. However, the singularity is
surrounded by the event horizon $r_{h}$, which is the positive
root of (the larger one if there are two positive real roots)

\begin{equation}
r^{2}-m-2 Q^{2} \ln r=0
\end{equation}

Thus if $r_{s}<r_{h}$ then the singularity will be covered by the
event horizon. While the singularity will be naked for $r_{s}\geq
r_{h}$. In this connection one may note that the event horizon is
independent of the coupling parameter $\alpha$.

\subsection{GEOMETRY OF THE BLACK HOLE THERMODYNAMICS IN EYMGB THEORY}

We discuss the geometry of thermodynamics of the black hole
described above. As the event horizon $r_{h}$ satisfies equation
(12) so we have,
\begin{equation}
m=r_{h}^{2}-2 Q^{2} \ln r_{h}
\end{equation}

The surface area of the event horizon is given by,
$A=2\pi^{2}r_{h}^{3}$
 and hence the entropy of the black
hole$[16]$ takes the form
\begin{equation}
S=\frac{K_{B}A}{4 G \hbar}=\frac{K_{B}\pi^{2}}{2 G \hbar}r_{h}^{3}
\end{equation}

Now choosing $\hbar=1$ and the Boltzmann constant appropriately,
we have, $S=r_{h}^{3}$

So $m$ can be obtained as a function of $S$ and $Q$ in the form
\begin{equation}
m=S^{\frac{2}{3}}-\frac{2}{3}Q^{2}\ln S
\end{equation}

From the energy conservation law of the black hole
\begin{equation}
dm=Tds+\phi dQ
\end{equation}

the temperature $(T)$ and electric potential$(\phi)$ of the black
hole are given by,
\begin{equation}
T=\left(\frac{\partial m}{\partial
S}\right)_{Q}=\frac{2}{3}S^{-\frac{1}{3}}-\frac{2 Q^{2}}{3 S}
\end{equation}
and
\begin{equation}
\phi=\left(\frac{\partial m}{\partial Q}\right)_{S}=-\frac{4
Q}{3}\ln S
\end{equation}

As the Weinhold metric is the Hessian of the internal energy (mass
parameter $'m'$ here) so its explicit form is
\begin{equation}
ds_{W}^{2}=\left[\frac{2}{3
S^{2}}\left(Q^{2}-\frac{1}{3}S^{\frac{2}{3}}\right)dS^{2}-\frac{8Q}{3S}dQdS-\frac{4}{3}\ln
S~ dQ^{2}\right]
\end{equation}

which can be diagonalized by the transformation
\begin{equation}
v=Q.u~~,~~u=\ln S
\end{equation}

and we obtain,
\begin{equation}
ds_{W}^{2}=\frac{4}{3}\left[\left\{\left(
\frac{1}{2}+\frac{1}{u}\right)\frac{v^{2}}{u^{2}}-\frac{1}{6}e^{\frac{2u}{3}}\right\}du^{2}-\frac{dv^{2}}{u}\right]
\end{equation}

which is curved and Lorentzian in nature provided
\begin{equation}
v^{2}>\frac{u^{3} e^{\frac{2 u}{3}}}{3(u+2)}
\end{equation}

Hence, using the conformal transformation $(3)$ the Ruppeiner
metric is given by,
\begin{equation}
ds_{R}^{2}=\frac{2e^{u}}{e^{\frac{2u}{3}}-\frac{v^{2}}{u^{2}}}\left[\left\{\left(
\frac{1}{2}+\frac{1}{u}\right)\frac{v^{2}}{u^{2}}-\frac{1}{6}e^{\frac{2u}{3}}\right\}du^{2}-\frac{dv^{2}}{u}\right]
\end{equation}

And the expression for the scalar curvature is,
\begin{equation}
R=-\frac{e^{-u}\left(3+u\right)\left(e^{2u}u^{6}(3+2u)-e^{\frac{4u}{3}}u^{4}
(-9+16u)v^{2}+3e^{\frac{2u}{3}}u^{2}(3+10u)v^{4}+27v^{6}\right)}{6\left(e^{\frac{2u}{3}}
u^{2}-v^{2}\right)\left(e^{\frac{2u}{3}}
u^{3}-3(2+u)v^{2}\right)^{2}}
\end{equation}

The region of $(u,v)$ is restricted by the inequality,
\begin{equation}
\frac{u e^{\frac{2 u}{3}}}{3(u+2)}<\frac{v^{2}}{u^{2}}<e^{\frac{2
u}{3}}
\end{equation}

The non-flatness of Ruppeiner metric indicates that the black hole
thermodynamics has statistical mechanical interactions. The
expression for the heat capacity with a fixed charge is given by,
\begin{equation}
c_{Q}=T\left(\frac{\partial S}{\partial T}\right)_{Q} =\frac{e^{u}
\left(e^{\frac{2 u}{3}}
-\frac{v^{2}}{u^{2}}\right)}{\left(\frac{v^{2}}{u^{2}}-\frac{1}{3}e^{\frac{2
u}{3}}\right)}
\end{equation}

Thus within the above admissible range$(24)$ $c_{Q}$ will be
positive provided
$\frac{v^{2}}{u^{2}}>\frac{1}{3}e^{\frac{2u}{3}}$ , while
$c_{Q}<0$ , if
$\frac{e^{\frac{2u}{3}}}{3(1+\frac{2}{u})}<\frac{v^{2}}{u^{2}}<\frac{1}{3}e^{\frac{2u}{3}}$.
So the EYMGB black hole will be unstable if $\frac{v}{u}\epsilon
\left(\frac{e^{\frac{u}{3}}}{\sqrt{3(1+\frac{2}{u})}}~,~\frac{e^{\frac{u}{3}}}{\sqrt{3}}\right)$and
it will be a stable one if
$\frac{u}{v}\epsilon\left(\frac{e^{\frac{u}{3}}}{\sqrt{3}}~,~e^{\frac{u}{3}}\right)$.
Further, at $\frac{v}{u}=e^{\frac{u}{3}}$, $c_{Q}$ changes sign
and the scalar curvature diverges. Therefore,
 there is a
 phase transition and corresponds to this critical point.

\section{GEOMETRY OF EMGB BLACK HOLE THERMODYNAMICS}

    This  section deals with  the thermodynamics of the black hole solution in
Einstein-Maxwell-Gauss-Bonnet theory with a comological constant.
At first the spherically symmetric black hole solution is
presented and then the thermodynamics of the black hole solution
has been examined geometrically in the parameter space.

\subsection{SPHERICALLY SYMMETRIC EINSTEIN-MAXWELL SOLUTION IN GAUSS-BONNET THEORY}

    The action in five dimensional space time $(M,~g_{\mu\nu})$
 that represents Einstein-Maxwell theory with a Gauss-Bonnet term and a
cosmological constant has the expression[17-19]
\begin{equation}
S=\frac{1}{2}\int_{M}
d^{5}x\sqrt{-g}[R-2\Lambda-\frac{1}{4}F_{\mu\nu}F^{\mu\nu} +\alpha
R_{GB}]
\end{equation}

where
$R_{GB}=R^{2}-4R_{\alpha\beta}R^{\alpha\beta}+R_{\alpha\beta\gamma\delta}R^{\alpha\beta\gamma\delta}$,
 is the Gauss-Bonnet term, $\alpha$ is the GB coupling parameter
having dimension $(length)^{2}$ ( $\alpha^{-1}$ is related to
string tension in heterotic super string theory), $\Lambda$ is the
cosmological constant and
$F_{\mu\nu}=\left(\partial_{\mu}A_{\nu}-\partial_{\nu}A_{\mu}\right)$
is the usual electromagnetic field tensor with $A_{\mu}$, the
vector potential.Now variation of this action with respect to the
metric tensor and $F_{\mu\nu}$gives the modified Einstein field
equations and Maxwell's equations
\begin{equation}
G_{\mu\nu}-\alpha H_{\mu\nu}+\Lambda g_{\mu\nu}=T_{\mu\nu}
\end{equation}

and
\begin{equation}
\nabla_{\mu}F_{\nu}^{\mu}=0
\end{equation}

where $H_{\mu\nu}$ is the Lovelock tensor (given by eq $(8)$) and
\begin{equation}
T_{\mu\nu}=2F^{\lambda}_{\mu}F_{\lambda\nu}-\frac{1}{2}F_{\lambda\sigma}F^{\lambda\sigma}g_{\mu\nu}
\end{equation}

is the electromagnetic stress tensor.

(Note that  the modified Einstein field equations (28) do not
contain any derivatives of the curvature terms and hence the field
equations remain second order).

 If the manifold $M$ is chosen to be five dimensional
spherically symmetric space-time having the line element
\begin{equation}
ds^{2}=-B(r)dt^{2}+B^{-1}(r)dr^{2}+r^{2}\left(d\theta_{1}^{2}+sin^{2}\theta_{1}(d\theta_{2}^{2}+\sin
^{2}\theta_{2}d\theta_{3}^{2})\right)
\end{equation}
with

$0 \leq \theta_{1}~~,~~\theta_{2}\leq \pi~~,~~0\leq\theta_{3}\leq
2\pi$,

 then solving the above field equations one obtains $[1, 19]$
\begin{equation}
B(r)=1+\frac{r^{2}}{4\alpha}-\frac{r^{2}}{4\alpha}\sqrt{1+\frac{16M\alpha}{\pi
r^{4}}-\frac{8Q^{2}\alpha}{3r^{6}}+\frac{4\Lambda\alpha}{3}}
\end{equation}

Here in an orthonormal frame the non-null components of the
electromagnetic tensors are $ F_{\hat{t}\hat{r}}=
-F_{\hat{r}\hat{t}} =\frac{Q}{4 \pi r^{3}}$. Note that in the
limit $\alpha \rightarrow 0$ one may recover the Einstein-Maxwell
solution with a cosmological constant. Further, in the limit with
$\Lambda=0$ we have the five-dimensional
Reissner-Nordstr$\ddot{o}$m Solution and hence the parameters
$M(>0)$ and $Q$ can be identified as the mass and charge
respectively. Moreover, for the solution $(32)$ to be well
defined, the radial coordinate $r$ must have a minimum value
($r_{min}$) so that the expression within the square root is
positive definite,i.e, the solution $(32)$ is well defined for
$r>r_{min}$ where $r_{min}$ satisfies

$$1+\frac{16m\alpha}{\pi
r_{min}^{4}}-\frac{8Q^{2}\alpha}{3r_{min}^{6}}+\frac{4\Lambda\alpha}{3}=0$$

The surface $r=(r_{min})$ corresponds to a curvature singularity.
However,depending on the values of the parameters this singular
surface may be surrounded by the event horizon (having radius
$r_{h}$ such that $B(r_{h})=0$)and the solution $(32)$ describes a
black hole solution known as EMGB black hole. On the other hand,
if no event horizon exists then the above solution represents a
naked singularity.

\subsection{GEOMETRIC APPROACH OF BLACK HOLE THERMODYNAMICS}

From the equation determining the event horizon for EMGB black
hole the mass parameter can be written as
\begin{equation}
M=\pi\alpha+\frac{\pi Q^{2}}{6}r_{h}^{-2}+\frac{\pi
r_{h}^{2}}{2}-\frac{\pi \Lambda}{12}r_{h}^{4}
\end{equation}

As before, with proper choice of units $S=r_{h}^3$ and hence
\begin{equation}
M=\pi\alpha+\frac{\pi Q^{2}}{6}S^{-\frac{2}{3}}+\frac{\pi
S^{\frac{2}{3}}}{2}-\frac{\pi \Lambda}{12}S^{\frac{4}{3}}
\end{equation}

 Thus from the first law of thermodynamics the temperature and electric
potential have the expressions
\begin{equation}
T=\frac{\pi}{3}S^{-\frac{1}{3}}-\frac{\pi}{9}Q^{2}S^{-\frac{5}{3}}-\frac{\pi
\Lambda}{9}S^{\frac{1}{3}}
\end{equation}

and
\begin{equation}
\phi=\frac{\pi Q}{3}S^{-\frac{2}{3}}
\end{equation}

The explicit expression for Weinhold metric is given by,
\begin{equation}
ds_{W}^{2}=\frac{\pi}{9
S^{\frac{5}{3}}}\left[-S^{\frac{1}{3}}\left(
1+\frac{\Lambda}{3}S^{\frac{2}{3}}-\frac{5Q^{2}}{3S^{\frac{4}{3}}}\right)dS^{2}-4QdQdS+3SdQ^{2}\right]
\end{equation}

It can be diagonalized to the form
\begin{equation}
ds_{W}^{2}=\frac{\pi}{9
S^{\frac{4}{3}}}\left[-\left\{\left(1-u^{2}\right)+\frac{\Lambda}{3}S^{\frac{2}{3}}\right\}dS^{2}+9S^{2}du^{2}\right]
\end{equation}

by the substitution
\begin{equation}
u=\frac{Q}{\surd3 S^{\frac{2}{3}}}
\end{equation}

The above Weinhold metric is non-flat and Lorentzian provided
\begin{equation}
u^2<1+\frac{\Lambda}{3}s^\frac{2}{3}
\end{equation}
Hence by the conformal transformation$(3)$ the expression for the
Ruppeiner metric is
\begin{equation}
dS_{R}^{2}=-\left\{\frac{\left(1-u^{2}\right)+\frac{\Lambda}{3}S^{\frac{2}{3}}}{\left(1-u^{2}\right)
-\frac{\Lambda}{3}S^{\frac{2}{3}}}\right\}
\frac{dS^{2}}{3S}+\frac{3Sdu^{2}}{\left\{\left(1-u^{2}\right)-\frac{\Lambda}{3}S^{\frac{2}{3}}\right\}}
\end{equation}

And the form  of the curvature scalar is given by,
\begin{equation}
R=\frac{\Lambda\left[5 \Lambda^{2}S^{\frac{4}{3}}+12\Lambda
S^{\frac{2}{3}}\left(u^{2}-2\right)+9\left(7u^{4}-2u^{2}-5\right)\right]}{3S^{\frac{1}{3}}\left(3+\Lambda
S^{\frac{2}{3}}-3y^{2}\right)^{2}\left(-3+\Lambda
S^{\frac{2}{3}}+3u^{2}\right)}
\end{equation}

 The curved nature of the Ruppeiner metric suggests that the
thermodynamics of the present black hole has statistical mechanics
analogue.

Now, for a given charge, the heat capaity has the expression
\begin{equation}
c_{Q}=3S\left(\frac{1-\frac{\Lambda}{3}S^{\frac{2}{3}}-u^{2}}{5u^{2}-1-\frac{\Lambda}{3}S^{\frac{2}{3}}}\right)
\end{equation}

Thus, $c_{Q}>0$ if

$$\frac{1}{5}\left(1+\frac{\Lambda}{3}S^{\frac{2}{3}}\right)<u^{2}<
\left(1-\frac{\Lambda}{3}S^{\frac{2}{3}}\right)$$
$$or$$
\begin{equation}
\left(1-\frac{\Lambda}{3}S^{\frac{2}{3}}\right)<u^{2}<\frac{1}{5}\left(1+\frac{\Lambda}{3}S^{\frac{2}{3}}\right)
\end{equation}

while $c_{Q}<0$ if

$$u^{2}<max\left\{\left(1-\frac{\Lambda}{3}S^{\frac{2}{3}}\right),
\frac{1}{5}\left(1+\frac{\Lambda}{3}S^{\frac{2}{3}}\right)
\right\}$$
$$or$$
\begin{equation}
u^{2}>min\left\{\left(1-\frac{\Lambda}{3}S^{\frac{2}{3}}\right),
\frac{1}{5}\left(1+\frac{\Lambda}{3}S^{\frac{2}{3}}\right)
\right\}
\end{equation}

Hence, $c_{Q}$ changes sign at
$u=\sqrt{\left(1-\frac{\Lambda}{3}S^{\frac{2}{3}}\right)}(=u_{1})$
and
$\sqrt{\frac{1}{5}\left(1+\frac{\Lambda}{3}S^\frac{2}{3}\right)}(=u_{2})$
but $c_{Q}$ diverges at $u_{2}$ while $c_{Q}=0$ at $u=u_{1}$. So
there is a possible phase transition at $u=u_{2}$. On the other
hand, the Ruppeiner metric coefficients change sign as $u$ crosses
$u_{1}$ but diverges as $u \rightarrow u_{1}$ while around $u=
u_{2}$ the metric coefficients are well behaved. Moreover, the
curvature scalar diverges at $u=u_{1}$. Therefore, there will be a
phase transition at $u=u_{1}$. Finally, note that if $\Lambda=0$
then the above result become very simple and the Ruppeiner metric
become flat. Also the results are in agreement with those of
[20].\\\\

{\bf Acknowledgement:}\\

A part of the work is done during a visit to IUCAA. The authors
are thankful to IUCAA for warm hospitality and
facilities of research. Ritabrata thanks Dr. Ujjal Debnath for valuable suggestions in preparing the manuscript.\\\\\\\\

$[1]$ S. W. Hawking , {\it Commun. Math. Phys.} {\bf43}, 199(1975).\\

$[2]$ J. D. Bekenstein, {\it Phys. Rev. D} {\bf 7}, 2333(1973).\\

$[3]$ J. M. Bardeen, B. Carter and S. W. Hawking, {\it Commun.
Math. Phys.} {\bf31}, 161(1973).\\

$[4]$ P. Hut, {\it Mon. Not. R. Astron Soc.} {\bf180}, 379(1977).\\

$[5]$. P. C. W. Davies, {\it Proc. Roy. Soc. Lond.~A} {\bf 353},
499(1977); {\it Rep. Prog. Phys.} {\bf41},1313 (1977); {\it Class.
Quant. Grav.} {\bf 6}, 1909(1989).\\

$[6]$ F. Weinhold, {\it J. Chem. Phy.} {\bf63}, 2479(1975).\\

$[7]$ G. Ruppeiner, {\it Rev. Mod. Phys.} {\bf67}, 605(1995);
{\bf68}, 313(E)(1996)\\

$[8]$ G. Ruppeiner, {\it Phys Rev. A} {\bf 20}, 1608(1979).\\

$[9]$ P. Candelas, G. T. Horowitz, A. Strominger and  E. Witten,
{\it Nucl. Phys. B}  {\bf 258},46(1985); M. B. Greens, J. H.
Schwarz and E. Witten, {\it "Superstring Theory"(Camb. Univ.
press, Cambridge, 1987)};
 J.Polchinski, {\it "Sting Theory"(Camb. Univ. press, Cambridge, 1998)}.\\

$[10]$ B. Zwiebach, {\it Phys. Lett. B} {\bf 156}, 315(1985);
 B. Zumino, {\it Phys. Rep.} {\bf 137}, 109(1986).\\

$[11]$ D. Lovelock, {\it J. Math. Phys.} {\bf12}, 498(1971).\\

$[12]$ C. Lanczos, {\it Ann. Math.} {\bf39}, 842(1938).\\

$[13]$ A. H. Chamseddine, {\it Phys. Lett.} {\bf B 233},
291(1989);
F.Muller-Hoissen, {\it Nucl. Phys.} {\bf B 349}, 235(1990).\\

$[14]$ M. Sami and N. Dadhich,{\it TSPU Vestnik} {\bf 44N7},25(2004)(arXiv:hep-th/0405016).\\

$[15]$ S. H. Mazharimousavi and M. Halisoy, {\it Phys. Rev.D} {\bf
76}, 087501(2007).\\

$[16]$ H. Falcke and F. W. Hehl (eds), {\it "The Galactic Black
Hole: Lectures on General Relativity and Astrophysics"(Bristol:
Institute of Physics Publishing, 2003)}.\\

$[17]$ D. G. Boulware and S. Deser, {\it Phys. Rev. Lett.} {\bf
55},2656(1985).\\

$[18]$ D. L. Wiltshire, {\it Phys. Letts. B.} {\bf B169},
36(1986);
{\it Phys. Rev. D} {\bf 38}, 2445(1988). \\

$[19]$ M. Thibeault, C. Simeone and E. F. Eirod, {\it Gen. Relt.
Grav.} {\bf38}, 1593(2006). \\

$[20]$ S. Chakraborty and T. Bandyopadhyay, {\it Class. Quant.
Grav.} {\bf25}, 245015(2008).

\end{document}